\documentclass[preprint]{revtex4-1}
\usepackage{graphicx}
\usepackage{dcolumn}
\usepackage{amssymb}
\usepackage{amsfonts}
\usepackage{amsbsy}
\usepackage{color}
\usepackage[english]{babel}
\usepackage{multirow}
\usepackage{natbib}
\begin{document}

\title{Testing a DBI model for the unification of dark matter and dark energy with Gamma-Ray Bursts.}

\author{A. Montiel and N. Bret\'on}
\affiliation{Dpto. de F\'isica, Centro de Investigaci\'on y de Estudios Avanzados del I. P. N.,\\ Av. IPN 2508, D.F., M\'exico}

\begin{abstract}
We study the range of consistency of a model based on a nonlinear scalar field Dirac-Born-Infeld action for the unification of dark matter and dark energy using Gamma-Ray Bursts at high-redshifts.
We use the sample of 59 high-redshift GRBs reported by Wei (2010), calibrated at low redshifts with the Union 2 sample of SNe Ia, thus avoiding the circularity problem. 
In this analysis, we also include the CMB7-year data and the baryonic acoustic peak BAO. Besides, it is calculated the parameter of the equation of state $w$, the deceleration parameter $q_0$ and the redshift of the transition to the decelerate-accelerated phase $z_t$. 
\end{abstract}
\maketitle

\section{Introduction}
The universe appears to consist of approximately 25$\%$ dark matter, which clusters and drives the formation of large-scale structure in the universe, and 70$\%$ dark energy, which drives the late-time acceleration of the universe. Since the nature of neither component is known with certainty, it is reasonable to ask whether a simpler model is possible, in which a single component acts as both dark matter and dark energy. In this work, we stick to this stream that considers DM and DE as different manifestations of the same component. 
Inspired in inflation models  \cite{Mukhanov}, recently it has been proposed to model dark energy with nonlinear scalar fields.
In particular, we address the model proposed in \cite{Ruth} that
describe the current acceleration with a Dirac-Born-Infeld (DBI) action: $S=- \int d^4x~a(t)^3 \left(f(\phi)^{-1}\left(\sqrt{1-f(\phi)\dot{\phi}^2}-1\right) + V(\phi)\right)$, characterized by two functions, the warp factor which should be positive, $f(\phi)>0$ and the inflation potential $V(\phi)$.
It has been shown in \cite{Ruth} that this model produces an expansionary cosmology mimicking a dark matter dominated background at early times and a dark energy dominated one at late times, therefore it is worth to 
update and extend the analysis presented in \cite{Ruth}.

We probe the model with the calibrated sample of GRBs  \cite{Montiel}. In the next section we introduced the main aspects of the model. Next we carry on the analysis with the observational data including SNe Ia, BAO for clusters with redshifts up approximately 2, as well as CMB5-year and CMB7-year from WMAP. Final remarks are draft in the last section

\section{The model with FRW spacetime}

The model considers a spatially flat Robertson-Walker geometry with a nonlinear scalar field of DBI type. It is assumed the customary perfect fluid interpretation with a barotropic EoS $p=(\Gamma -1) \rho$, then $\rho=\frac{\gamma-1}{f}+V(\phi)\label{rhodef}$ and $p=\frac{\gamma-1}{\gamma f}-V(\phi)$ with $\gamma=\frac{1}{\sqrt{1- f(\phi)\dot\phi^2}}$ where $f$ and $V$ are in general arbitrary functions. We shall consider the simplest case in which both $f=f_0$ and $V=V_0$ are constant. The goal is to obtain a purely kinetic model so that the field $\phi$ depends solely on the scale factor and as a consequence the same holds for the effective pressure and energy density. Using the conservation equation, $\dot{\rho}+3H\Gamma \rho=0$ it can be derived the scaling between the scalar field and the expansion factor $a(t)$: $\dot\phi=\frac{c}{\gamma}\left(\frac{a_0}{a}\right)^3$. With this last expression it is possible to write
\begin{equation}
\gamma^2=1+c^2f_0\left(\frac{a_0}{a}\right)^6,\label{gsq}
\end{equation}
with  $c$ an arbitrary integration constant and  $a_0$ the value of the scale factor today, which we fix as $a_0=1$.
From previous equation, it can be seen that at very late times, $a\gg a_0$, $\gamma \simeq 1$ $\Rightarrow$ $\rho \sim V_0$, whereas at early epochs, $a \ll a_0$, $\gamma \simeq a^{-3}$ $\Rightarrow$ $\rho \sim 1/a^3$.
Thus, the solution interpolates between a dust and a de Sitter model, with $V_0$ acting as a cosmological constant and no trace left of the genuinely DBI degree of freedom $f_0$.

On the other hand, the Friedmann equation turns out to be
\begin{equation}
3H^2=f_0^{-1}\left({\sqrt{1+c^2f_0\left(\frac{a_0}{a}\right)^6}-1}\right)+V_0+\rho_{r0}\left(\frac{a_0}{a}\right)^4.
\end{equation}
where $\rho_{r0}$ is the radiation energy density;
in terms of the fractional energy densities and the redshift 
\begin{equation}
\frac{H^2}{H_0^2}=\sqrt{\Omega_f^2+\Omega_c^2(1+z)^6}+\Omega_{\Lambda}+\Omega_r(1+z)^4,
\label{eq:fri}
\end{equation}

where $\Omega_f=\frac{1}{3H_0^2f_0}$, $\Omega_c=\frac{c}{3H_0^2\sqrt{f_0}}$, $\Omega_{\Lambda}=\frac{f_0V_0-1}{3H_0^2f_0}$, $\Omega_{r}=\frac{\rho_{r0}}{3H_0^2}$. 

In addition, it is possible investigate the redshift dependence of the parameter of the EoS $w(z)=p/ \rho$, and the deceleration parameter which can be expressed as  $q(z)=(3(1-\Omega_c(1+z)^3/H^2)w(z)+1)/2$. To obtain the transition redshift we just compute $q(z_t)=0$.

\section{Observational Data Analysis and Results}

Since the earliest evidence of tight correlations in gamma-ray bursts spectral properties, the possibility arose of using GRBs as standard candles. Being so GRBs may open a window in redshift as far as $z \sim 8$, extending then the attainable range provided by SNe Ia observations. 

The updated distance moduli of the 59 GRBs at $z>1.4$ are obtained through the Amati relation that connects $E_p=E_{p, obs}(1+z)$ with the isotropic equivalent energy $E_{iso}=4\pi d^2_L S_{bolo} (1+z)^{-1}$, where $S_{bolo}$ is the bolometric fluence of gamma rays in the GRB at redshift $z$ and $d_L$ is the luminosity distance of the GRB (of the hosting galaxy) \cite{Amati}. The Amati relation is calibrated from 50 GRBs at $z<1.4$ following the cosmology-independent calibration method proposed by Liang (2008) \cite{Liang} using the Union2 set of 557 SNe Ia. For details of the calibration see \cite{Montiel}. We also consider the shift parameter set $(l_A, R, z_\ast)$ from 5-year WMAP and 7-year WMAP \cite{wmap7} and the BAO parameter set $\textbf{v}=\left\{\frac{r_s(z_{\rm{drag}})}{D_V(0.2)},\frac{r_s(z_{\rm{drag}})}{D_V(0.35)}\right\}$ \cite{Percival9}. These combination of test allows to take into account the early, mid and late time behavior of the model.

We calculate the \textit{best estimated values} for the parameters $\Omega_f$ and $\Omega_c$ and the \textit{goodness-of-fit} of the model to the data by $\chi^2$-minimization and then compute the confidence intervals for ($\Omega_f$, $\Omega_c$) to constrain their possible values.

The best-fit value for the DBI model parameters ($\Omega_c$, $\Omega_f$), with 1-$\sigma$ uncertainties, $\chi^2_{d.o.f.}$, as well as $w_0$, $z_t$ and $q_0$ using SNe Ia + GRBs, SNe Ia + CMB7 + BAO and  SNe Ia + GRBs + CMB7 + BAO are summarized in Table \ref{tabla1}. The results with SNe Ia + CMB5 + BAO and  SNe Ia + GRBs + CMB5 + BAO are very similar to those obtained with CMB7-year from WMAP. 

\begin{table*}
    \begin{tabular}{lccc}
\hline
& SNe Ia + GRBs 
  & SNe Ia + CMB7 + BAO
  &SNe Ia + GRBs + CMB7 + BAO  \\
\hline
$\Omega_c$ & 0.3002$^{+ 0.0498}_{-0.0476}$& 0.2549$^{+ 0.0112}_{-0.0068}$& 0.2553$^{+0.0115 }_{-0.0069}$ \\ 
$\Omega_f$ & 0.3106$^{+ 0.1894}_{-0.3106}$& 0.0918$^{+ 0.1735}_{-0.0918}$& 0.0968$^{+0.1746 }_{-0.0968}$\\ 
$w_0$      &-1.1307$^{+0.1310}_{-0.0775}$&-1.0201$^{+0.0202}_{-0.0849}$& -1.0221$^{+0.0223 }_{-0.0857}$\\ 
$z_t$      &-1.1307$^{+0.1310}_{-0.0775}$&0.7936 $^{+0.0226}_{-0.0485}$& 0.7916$^{+0.0230}_{-0.0493}$\\ 
$q_0$      &-0.6869 $^{+0.1588}_{-0.1097}$&-0.6401 $^{+0.0287}_{-0.0954}$& -0.6418 $^{+0.0308}_{-0.0963}$\\ 
$\chi^2_{d.o.f.}$&0.9472& 1.0062   & 0.9740\\
\hline
\end{tabular}
\caption{The best-fit value for the parameters of the  model based in a DBI action ($\Omega_c$, $\Omega_f$) with 1-$\sigma$ uncertainties, $\chi^2_{d.o.f.}$, as well as $w_0$, $z_t$ and $q_0$ using SNe Ia + GRBs, SNe Ia + CMB7 + BAO and SNe Ia + GRBs + CMB7 + BAO.  The subscript "t" stands for the "transition" and the subscript "d.o.f", stands for "degrees of freedom". The confidence intervals are shown in figure \ref{fig2}.}
\label{tabla1}
  \end{table*}

\begin{figure*}
  \includegraphics[height=.25\textheight]{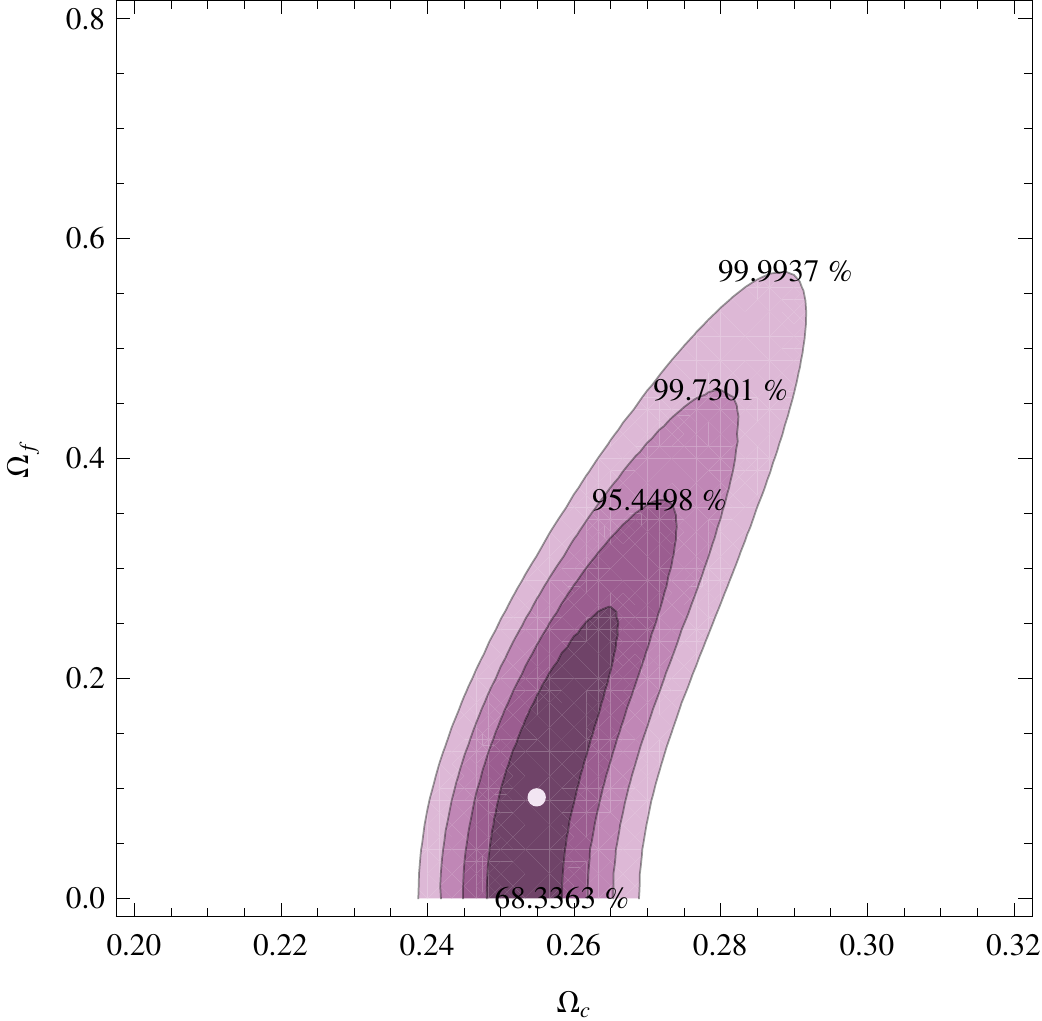}~~~~~~~~~~~~~~~~~~~~~~~~~
  \includegraphics[height=.25\textheight]{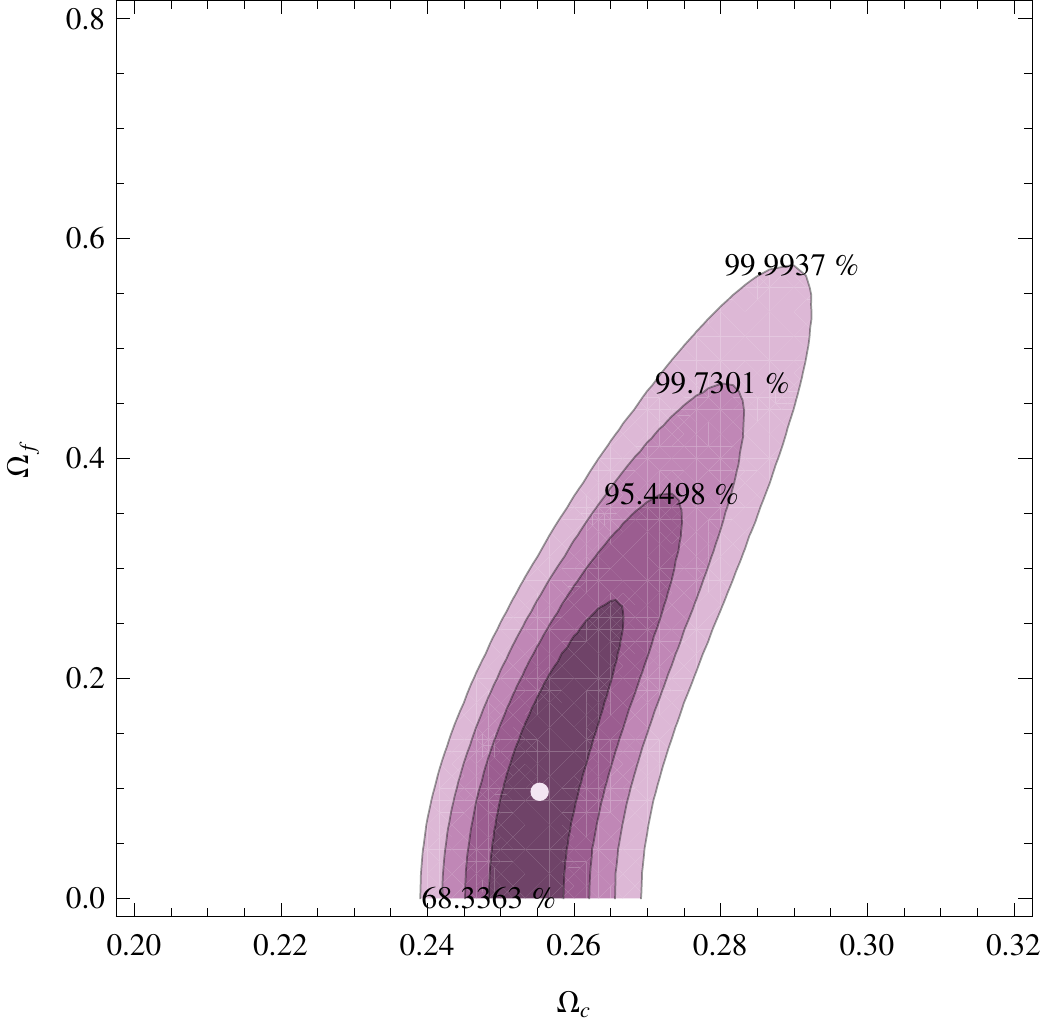}
  \caption{The joint confidence regions in the ($\Omega_c$, $\Omega_f$) plane for the DBI model. The contours correspond to 1$\sigma$-4$\sigma$ confidence regions using SNe Ia + CMB7 + BAO (left panel) and SNe Ia + GRBs + CMB7 + BAO (right panel). }
  \label{fig2}
\end{figure*}

\section{Conclusions}

In this work  we use  the updated sample of 109 
calibrated GRBs, avoiding the well known circularity problem, in order to 
test a cosmological model. In particular, we studied a model based on a 
nonlinear scalar field DBI action proposed by Chimento-Lazkoz-Sendra 
(2010) for the unification of dark matter and dark energy. Probing with 
557 SNe Ia, 59 GRBs (at high redshifts), CMB7-year data from WMAP and BAO 
we obtained as best-fit model parameters ($\Omega_c= 
0.2553^{+0.0115}_{-0.0069}$, $\Omega_f= 0.0968^{+0.1746}_{-0.0968}$). 
Our results are in 
good agreement with previous conclusions in \cite{Ruth}, namely, the 
model is better suited to the observations than the Chaplygin gas.
Including GRBs the obtained $w=-1.1307_{-0.0775}^{+0.1310}$ points to a 
phantom behaviour 
in a stronger way than the previous analysis in \cite{Ruth}; however
at the present precision $\Lambda$CDM cannot be ruled out.
On the other hand, 
from the analysis of the deceleration parameter we obtained an accelerated 
epoch with a transition from decelerated-accelerated epoch at 
$z_t=0.7916^{+0.0230}_{-0.0493}$; as for the deceleration parameter the 
obtained value when including GRBs,
$q_0=-0.6869^{+0.1588}_{-0.1097}$, is closer to the accepted value.

The obtained results confirm GRBs data 
as a valuable tool in probing models where dark matter and dark energy are 
unified.


\begin{acknowledgments}
A. M acknowledges a Ph D grant (65902) by Conacyt, M\'exico. 
\end{acknowledgments}

\end{document}